\documentclass[prb,aps,twocolumn,showpacs]{revtex4}

\usepackage{amsfonts,amsmath,bm}
\usepackage[OT4]{fontenc}
\usepackage{mathtools}
\usepackage{graphicx}
\usepackage{epstopdf}

\newcommand{\bk}{b_{\bm{k}}}

\newcommand{\bmkd}{b_{-\bm{k}}^{\dag}}
\newcommand{\bkd}{b_{\bm{k}}^{\dag}}

\newcommand{\sumk}{\sum_{\bm{k}}}

\newcommand{\omk}{\omega_{\bm{k}}}

\newcommand{\ket}{\rangle}

\newcommand{\rl}{\rangle\!\langle}

\DeclarePairedDelimiter\norm{\lVert}{\rVert}%
\DeclarePairedDelimiter\abs{\lvert}{\rvert}%
\makeatletter
\let\oldabs\abs
\def\abs{\@ifstar{\oldabs}{\oldabs*}}
\let\oldnorm\norm
\def\norm{\@ifstar{\oldnorm}{\oldnorm*}}
\makeatother
\begin{document}

\title{Decay and persistence of spatial coherence during phonon-assisted relaxation in
double quantum dots}  
\author{Pawe{\l} Karwat}
\email{Pawel.Karwat@pwr.edu.pl}
\affiliation{Department of Theoretical Physics, Wroc{\l}aw University of
Technology, 50-370 Wroc{\l}aw, Poland}
\author{Pawe{\l} Machnikowski} 
\affiliation{Department of Theoretical Physics, Wroc{\l}aw University of
Technology, 50-370 Wroc{\l}aw, Poland}

\begin{abstract}
We present a theoretical study of the evolution of spatial coherence during intraband
relaxation between exciton states in a pair of vertically stacked semiconductor quantum
dots coupled to acoustic phonons. We show that spatial coherence can be transferred to the
ground state even in a system of uncoupled non-identical quantum dots if a particular kind
of degeneracy between the inter-level energy splittings is present. The phonon-assisted
mechanism of coherence transfer leads to a dependence of the amount of the resulting
coherence on the inter-dot distance and temperature. We analyze also the impact of
carrier-phonon dynamics on a coupled system, where spatial coherence is present in the
delocalized ground state.
\end{abstract}


\maketitle
\section{Introduction}
\label{sec:intro}
The physical properties of systems consisting of two or more semiconductor quantum dots
(QDs) are rich and cannot be reduced to single-dot characteristics. Experiments
demonstrate that collective emission effects play a role in the optical response of such
systems \cite{scheibner07,mazur10b}, which can be attributed to the formation of
delocalized bright and dark states in the presence of inter-dot coupling, even in
energetically inhomogeneous structures \cite{sitek07a}. Phonon-induced carrier transfer
\cite{nakaoka06} in double quantum dots (DQDs) has been predicted to affect the optical
emission by redistributing the occupations between bright and dark states \cite{karwat11}.
Even though basic quantum-optical concepts of collective emission or superradiance can be
transferred from atomic systems to QDs, the relaxation and dephasing effects induced by
the coupling to the lattice environment are specific to QD systems and require an extended
treatment, as compared to the theory developed in the atomic context \cite{gross82}.\\
\indent The experimental evidence of collective effects in the emission from
self-assembled QD samples, both closely spaced stacks \cite{mazur10b} as well as natural
planar ensembles \cite{scheibner07} is particularly interesting, as it suggests that such
systems cannot be treated as ensembles of independent emitters. The enhanced,
``superradiant'' emission from the ground state of a QD ensemble is rather intriguing in
view of the huge energy inhomogeneity of such a system (as compared to the emission line
width). While the optical effect has been successfully modeled under the assumption that
the QDs are coupled not only by the relatively weak long-range dipole interactions but
also by short range ones (which might result from a combination of Coulomb couplings and
tunneling)\cite{kozub12}, the observed difference\cite{scheibner07} between the
time-resolved ensemble emission under quasi-resonant excitation (optical transition to
higher confined shells, leading to a collective enhancement of emission) and non-resonant
excitation (transition to wetting layer or bulk states with no enhancement of emission
observed) has not been discussed. Since collective emission relies on constructive
interference of emission amplitudes from different atoms or QDs, which is possible due to
the formation of coherently delocalized Dicke states, description of the dynamics of
spatial coherences in coupled QD systems seems to be crucial for the general understanding
of the optical properties of such systems. In particular, since spatial coherence can
appear as a result of quasi-resonant coherent optical excitation, it is interesting to
find out if, and under what conditions, it can be transferred to the ground state
manifold, from which the emission takes place.\\
\indent The principal mechanism of intraband transitions in self-assembled QDs is the
carrier-phonon coupling and, indeed, models based on phonon-induced processes successfully
reproduce experimental data on carrier relaxation\cite{zibik09}. While in single dots
phonon-related effects essentially include relaxation\cite{verzelen02b} and pure
dephasing\cite{krummheuer02}, in DQDs one deals with a broader variety of processes which
is due to the rich structure of carrier states in these systems. Apart from intra-dot
relaxation, these processes may include inter-dot phonon-assisted
tunneling\cite{gawarecki10,gawarecki12a,wijesundara11,nakaoka04,muller12} and
Coulomb-mediated transfer\cite{govorov03,rozbicki08a,gerardot05} of carriers, as well as
dephasing of spatial coherence of carrier states delocalized over two
dots\cite{muljarov05,machnikowski06d}.\\
\indent In this paper, we study theoretically another phonon-related effect in a DQD
structure, which may be of importance for the collective luminescence: the phonon-assisted
transfer of spatial coherence from the excited shells of the DQD system to the ground
state manifold (corresponding to the quasi-resonant excitation conditions in an optical
experiment). We show that after a coherent excitation of the DQD to a spatially
delocalized excited states (which is a reasonable assumption in view of the small DQD size
as compared to the relevant wavelength of the laser light), the spatial coherence can
indeed be conserved during phonon-assisted relaxation in a system of decoupled QDs
provided that the inter-level energy differences in the two dots do not differ
considerably. While this effect is captured by a Markovian model of the system dynamics,
it is lost in the secular approximation inherent in the most standard Lindblad equation
approach. We discuss also the evolution of the spatial coherence in a system of coupled
QDs (a quantum dot molecule), where a certain degree of coherent delocalization is built
in already in the system ground state and various dynamical scenarios are possible
depending on the ratio between the intra-dot and inter-dot relaxation rates.\\
\indent The paper is organized as follows. In Sec.~\ref{sec:model}, we define the model
under study. Next, in Sec.~\ref{sec:results} we present the simulations of the evolution
of the spatial coherence in DQDs. In particular, Sec.~\ref{subsec:uncoupled} deals with
uncoupled QDs, while Sec.~\ref{subsec:coupled} describes coupled system.
Sec.~\ref{sec:conclusions} contains the final discussion and concludes the paper.

\section{Model}
\label{sec:model}
\begin{figure}[t]
\begin{center}
\includegraphics[width=70mm]{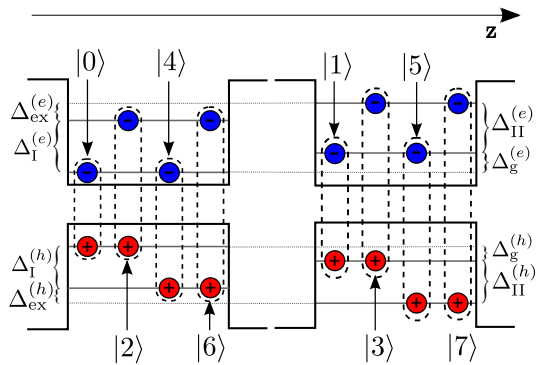}
\end{center}
\caption{\label{fig:system}(Color online) The basis exciton states in the system,
$\Delta_{\mathrm{I/II}}^{(e/h)}$ and $\Delta_{\mathrm{g/ex}}^{(e/h)}$are the parameters
describing the transition energy mismatch in the dots for electron and the hole,
respectively.}
\end{figure}

The system under study is made up of two large self-assembled and vertically stacked QDs
\cite{xie95} occupied by a single exciton, coupled to longitudinal acoustic phonons via
the deformation potential. We assume that the differences between the corresponding
single-particle levels in the two dots are in the range of a few meV, hence much smaller
than the energy of the electron-hole Coulomb attraction. The lowest exciton eigenstates
are then formed by spatially direct states, that is, configurations in which the
electron-hole pairs reside in the same dot \cite{szafran05,szafran08} (presented in
Fig.~\ref{fig:system}). We also assume that the spins of the carriers are fixed. Under
these assumptions, the Hilbert space is spanned by eight basis exciton states
shown~in~Fig.~\ref{fig:system}. Here the relative energies of the exciton states are
parametrized by $\Delta_{\mathrm{I/II}}^{(e/h)}, \Delta_{\mathrm{g/ex}}^{(e/h)}$ (see
Fig.~\ref{fig:system}). For each particle, three out of the four parameters are
independent. For simplicity, we assume that the electron-hole Coulomb energy is similar in
all the spatially direct configurations.

The total Hamiltonian of the system is
\begin{equation*}
H=H_{\mathrm{DQD}}
+H_{\mathrm{ph}}+H_{\mathrm{c-ph}}.
\end{equation*}
The first term describes exciton states in the DQD structure and has
the form
\begin{eqnarray}
H_{\mathrm{DQD}} & = &   \sum_{m}\epsilon_{m}|m\rl m|
+\sum_{m~\mathrm{even}}V_{m}\left(|m\rl m+1|+\mathrm{h.c.}\right),\nonumber 
\label{ham-DQD-S}
\end{eqnarray}
where $\epsilon_{m}$ are the exciton energies, and the inter-dot couplings are
parametrized by $V_{m}$ which is assumed real. We consider the simplest model that is able
to account for the effect in question and take into account only a single excited level
for each carrier in each dot. We include only couplings between states belonging to the
same electron and hole shell, which are energetically close to each other. Even and odd
labels correspond to the lower and upper dot, respectively. In the simplest approach, the
electron and hole wave functions in the dots are modeled by
identical anisotropic Gaussians 
with identical extensions $l$ in the $xy$ plane
and $l_{z}$ along the growth axis $z$ for both particles,
\begin{equation}\label{wf_g}
\psi_{\mathrm{g}}(\bm{r})\sim\exp\left[ 
-\frac{r^{2}_{\perp}}{2l^{2}} - \frac{(z\pm D/2)^{2}}{2l_{z}^{2}}
 \right],
\end{equation}
\begin{equation}\label{wf_ex}
\psi_{\mathrm{ex}}(\bm{r})\sim r_{\perp}\psi_{\mathrm{g}}(\bm{r}) e^{i\varphi},
\end{equation}
where $r_{\perp} = \sqrt{x^{2}+y^{2}}$ and $\varphi$ are cylindrical coordinates in the xy
plane, the g/ex indices refer to the ground and first excited level, and $D$ is the
distance between the dots. The exciton wave functions are assumed in the form of products
of single-particle functions. 

We assume that inter-dot carrier-phonon couplings  are negligible due to small overlap of
the wave functions confined in different dots. Hence, for the interaction of carriers
confined in the
DQD with phonons we only retain intra-dot single-particle terms (thus neglecting also the
contribution from Coulomb correlations \cite{daniels13}). The exciton-phonon coupling is
then modeled by the Hamiltonian
\begin{eqnarray}\label{ham-phon}
H_{\mathrm{c-ph}}&=&\sum_{(mn)}
|m\rl n|\sumk F_{mn}(\bm{k})(\bk+\bmkd)\nonumber\\
&+& \sum_{n}
|n\rl n|\sumk F_{nn}(\bm{k})(\bk+\bmkd),
\end{eqnarray}
where $(mn)$ denotes summation over pairs of exciton states related by single-particle
relaxation within a single QD and $\bk,\bkd$ are the bosonic operators of the phonon
modes. For uncorrelated (product) exciton wave functions, the off-diagonal coupling
constants for the exciton states are equal to the electron or hole coupling constant,
depending on which particle is involved in the $m\leftrightarrow n$ transition (see
Fig.~\ref{fig:system}), and have the form\cite{grodecka05a},
\begin{equation}
\label{ff}
F_{mn}(\bm{k})=F_{e/h}(\bm{k}) e^{\pm ik_{z}D/2},
\end{equation}  
where 
\begin{displaymath}
F_{e/h}(\bm{k})=\sigma_{\mathrm{e/h}}\sqrt{\frac{k}{2\varrho vc_{l}}}i\frac{k_{\perp}l}{2}
\exp\left[
-\frac{l_{z}^{2}k_{z}^{2}+l^{2}k_{\bot}^{2}}{4}\right]
e^{i\phi}
\end{displaymath}
and the ``+/-" sign corresponds to the lower and upper QD. 
The diagonal coupling constants are
\begin{eqnarray}
\lefteqn{F_{nn}(\bm{k})=} \nonumber\\
& &
\left[\sigma_{\mathrm{e}}\left(1-\xi_{n}^{(e)}\frac{k^{2}_{\perp}l^{2}}{4}\right)-\sigma_{
\mathrm{h}}\left(1-\xi_{n}^{(h)}\frac{k^{2}_{\perp}l^{2}}{4}\right)\right]\nonumber\\
&\quad&\times\sqrt{\frac{k}{2\varrho
vc_{l}}}\exp\left[-\frac{l_{z}^{2}k_{z}^{2}+l^{2}k_{\bot}^{2}}{4}\right], \nonumber
\end{eqnarray} 
where $\xi_{n}^{(e/h)} =1$ if the electron/hole is in an excited state in the exciton
state $n$ and $\xi_{n}^{(e/h)} =0$ otherwise.
Here $v$ is the normalization volume,  
$k_{\bot/z}$ are momentum
components in the $xy$ plane and along the $z$ axis, $\phi$ is the corresponding azimuthal
angle, $\sigma_{\mathrm{e/h}}$ are deformation potential constants for
electrons/holes, $c_{l}$ is the speed of longitudinal sound,
and $\varrho$ is the crystal density. 

The phonon modes are described by the free phonon Hamiltonian
\begin{displaymath}
H_{\mathrm{ph}}=\sumk\hbar\omk\bkd\bk,
\end{displaymath}
where $\omk$ are the corresponding
frequencies. We assume a linear dispersion relation for phonons.

In our numerical simulations,
we take the parameters corresponding to large, flat self-assembled 
InAs/GaAs QDs: $\sigma_{\mathrm{e}}=7$~eV, $\sigma_{\mathrm{h}}=-3.5$~eV,
$\rho=5350$~kg/m$^{3}$, $c_{l}=5150$~m/s, 
the wave function parameters
$l=20$~nm, $l_{z}=1$~nm.

\section{Simulation method and results}
\label{sec:results}
\subsection{Uncoupled QDs}
\label{subsec:uncoupled}
\subsubsection*{Evolution equations}
The general Master equation in the Markov limit and in the interaction picture
\cite{breuer02} has the form
\begin{eqnarray}
\dot{\rho}&=&\pi\sum_{klmn}e^{i(\omega_{kl}-\omega_{mn})t}R_{lkmn}(\omega_{mn})\nonumber
\\
&\quad&\times \big\lbrack|m\rl n|\rho|l\rl k|-|l\rl k|m\rl n|\rho\big\rbrack +
\mathrm{h.c.},
\label{master_eq}
\end{eqnarray}
where 
\begin{eqnarray}
R_{lkmn}(\omega)&=&\frac{1}{\hbar^{2}}\sum_{\bm{k}}F_{lk}(\bm{k})F_{nm}^{*}(\bm{k})\nonumber\\
&\times& \delta(|\omega|-\omega_{\bm{k}})|n_{B}(\omega)+1|.
\label{spectral_density}
\end{eqnarray}
Note that $R_{lknm}(\omega) = R_{mnkl}(\omega) = R^{*}_{nmlk}(\omega)$. This equation can
be transformed to the Lindblad form (see Appendix A) and therefore preserves the
physicality (in particular, positive definiteness) of the density matrix. Here
$\omega_{mn} = (\epsilon_{n}-\epsilon_{m})/\hbar$ and $n_{\mathrm{B}}(\omega)$ is the Bose
distribution function. For the phonon-induced intraband transitions, the frequencies
$\omega_{mn}$ are determined by the parameters $\Delta^{(e/h)}_{\mathrm{I/II}},
\Delta^{(e/h)}_{\mathrm{g/ex}}$ (see Fig.~\ref{fig:system}). In deriving
Eq.~\eqref{master_eq}, we have assumed that the reservoir memory is short compared to the
time scales of the system evolution and restricted the description to times long compared
to the memory time \cite{breuer02}. However, no secular approximation has been performed
at this point. 

\begin{figure}[tb]
\includegraphics[width=80mm]{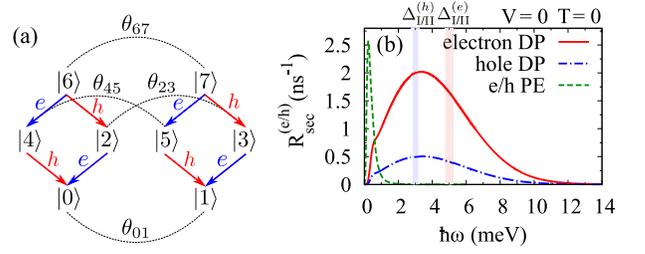}
\caption{\label{fig:sd_DP_PE}(Color online) (a) The possible transitions between the
states for the electron and the hole in the system. Red and blue arrows indicate
transitions between exciton states involving electron and hole relaxation, respectively,
while black dotted arrows show tunnel couplings between the states localized in different
dots. (b) The secular phonon spectral densities  at $T = 0$~K, for an electron transition
(red solid line), and hole transition (blue dashed line) respectively, coupled via
deformation potential and for the electron/hole (green dashed line) coupled via
piezoelectric field. The vertical stripes correspond to the range of changes of the
energy-level differences in each dot ($\Delta^{(e)}_{\mathrm{I/II}}$ and
$\Delta^{(h)}_{\mathrm{I/II}}$), for an electron and a hole, respectively.}
\end{figure}

In the case of no degeneracy between intraband transition energies
($\Delta^{(e)}_{\mathrm{I}}$ and $\Delta^{(h)}_{\mathrm{I}}$ sufficiently different from
$\Delta^{(e)}_{\mathrm{II}}$ and $\Delta^{(h)}_{\mathrm{II}}$, respectively), $\omega_{kl}
- \omega_{mn}$ is large unless $k = m$, $l = n$, hence one can apply the secular
approximation, neglecting terms with $k \neq m$ or $l \neq n$ and get a Lindblad equation
in the form \cite{breuer02}
\begin{eqnarray}
\label{lindblad_eq}
    \lefteqn{\dot{{\rho}} =} \nonumber\\
    & & \pi\sum_{mn}R_{nmmn}(\omega_{mn})\Big\lbrack|m\rl n|\rho |n\rl m|\!-\!|l\rl k|m\rl
n|\rho  \Big\rbrack + \mathrm{h.c.} \nonumber\\
    & &  = 2\pi\!\sum_{mn}\!R_{nmmn}(\omega_{mn})\Big\lbrack|m\rl n|\rho|n\rl
m|\!-\!\frac{1}{2}\big\lbrace |n\rl n|, \rho \big\rbrace \Big\rbrack
\end{eqnarray}
where $R_{nmmn}(\omega_{mn})$ is the secular spectral density describing the
phonon-assisted relaxation of either an electron (e.g. $R_{2002}(\omega_{02})$) or a hole
(e.g. $R_{4004}(\omega_{04})$) within one dot, according to the electron or hole character
of $F_{mn}(\bm{k})$. Thus, we only have two different secular spectral densities, denoted
$R^{(e)}_{\mathrm{sec}}, R^{(h)}_{\mathrm{sec}}$. Both of them are presented in
Fig.~\ref{fig:sd_DP_PE} with red solid and blue dash-dotted lines, respectively. They do
not depend on the distance between the dots. For comparison, we plot in
Fig.~\ref{fig:sd_DP_PE} also the spectral density for piezoelectric phonons, calculated
using the standard model for piezoelectric carrier-phonon couplings\cite{wu05} (green
dashed line). As can be seen, this coupling contributes only at very low frequencies and
can be neglected in view of much larger energy level spacings assumed in our model
(vertical stripes in Fig.~\ref{fig:sd_DP_PE}).

\begin{figure}[tb]
\includegraphics[width=85mm]{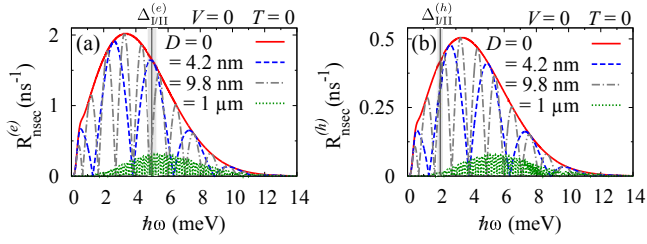}
\caption{\label{fig:sd_V0}(Color online) The non-secular phonon spectral densities at $T =
0$~K for uncoupled QDs, for an electron transition (a), and hole transition (b). Different
lines correspond to the three values of the inter-dot distance, as shown. Here, the
vertical gray lines correspond to the values of the energy-level differences used in our
simulations in Sec.~\ref{subsec:uncoupled} ($\Delta^{(e)}_{\mathrm{I/II}} = 5$~meV and
$\Delta^{(h)}_{\mathrm{I/II}} = 2$~meV, for the electron and the hole, respectively).}
\end{figure}

From Eq.~\eqref{lindblad_eq} we get for the inter-dot coherence in the ground state
\begin{equation}
\frac{d}{dt}\langle 0|\rho|1\rangle = -\frac{1}{2}\sum_{m}(\Gamma_{0m}+\Gamma_{1m})\langle
0|\rho|1\rangle,
\end{equation} 
where $\Gamma_{nm} = 2\pi R_{nmmn}(\omega_{mn})$.
In this case we have a purely exponential decay and no coherence can appear dynamically.\\
\indent Now, let us assume that $\omega_{kl}-\omega_{mn}$ can be small for pairs of
transitions $(kl)$ and $(mn)$ corresponding to relaxation of a given particle (electron or
hole) in different dots. Consistently with this, let us keep the corresponding terms in
Eq.~\eqref{master_eq}. For example, for $(mn)=(02)$ the terms with $(kl)=(02)$ and
$(kl)=(13)$ are kept (both corresponding to electron relaxation in the two dots with the
hole in the ground shell). According to Fig.~\ref{fig:system}, this condition corresponds
to the degeneracy of intraband transition energies $\Delta_{\mathrm{I}}^{(e/h)} \approx
\Delta_{\mathrm{II}}^{(e/h)}$.
Now the equation of motion for the ground-state spatial coherence of interest is

\begin{eqnarray}
\label{master_eq2}
\lefteqn{\frac{d}{dt}\langle 0|\rho|1\rangle=} \\
&\quad&-\frac{1}{2}\sum_{m}(\Gamma_{0m}+\Gamma_{1m})\langle 0|\rho|1\rangle \nonumber\\
&\quad& + \pi\sum_{ln}\Big\lbrack R_{l10n}(\omega_{0n})+R_{1ln0}(\omega_{1l})\Big\rbrack
e^{i(\omega_{1l}-\omega_{0n})t}\langle n|\rho|l\rangle,\nonumber
\end{eqnarray}
where $(l,n) = (3,2)$ or $(5,4)$, corresponding to electron and hole relaxation,
respectively. For the geometry chosen in our model, the non-secular spectral densities
$R_{l10n}(\omega)$ are real and again we have only two different non-secular spectral
densities, depending on whether the transitions $l \rightarrow 1$ and $n \rightarrow 0$
involve an electron or hole relaxation. They will be denoted by
$R^{(e/h)}_{\mathrm{nsec}}$. For this kind of a non-secular spectral density, the
expression in Eq.~\eqref{spectral_density} contains a factor $\cos(k_{z}D)$ (originating
from the phase factor in Eq.~\eqref{ff}) depending on the distance between the dots, 
\begin{eqnarray}
\label{R_D}
\mathrm{R}^{(e/h)}_{\mathrm{nsec}}(\omega)&=&\frac{1}{\hbar^{2}}\sum_{\bm{k}}|F_{e/h}(\bm{
k})|^{2}\cos{(k_zD)} \\
&\quad&\times\delta(|\omega|-\omega_{\bm{k}})|n_{B}(\omega)+1|.\nonumber
\end{eqnarray}
This oscillating factor in the integrand leads to oscillations in the spectral densities
as shown in Fig.~\ref{fig:sd_V0}.
\begin{figure}[bt]
\includegraphics[width=85mm]{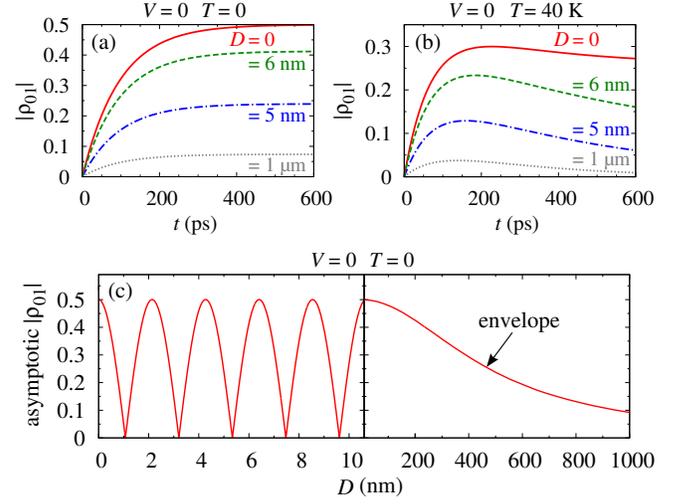}
\caption{\label{fig:identical_T0_40}(Color online) (a,b) The evolution of the spatial
coherence for uncoupled QDs in the resonant case ($\Delta^{(e)}_{\mathrm{I/II}} = 5$~meV
and $\Delta^{(h)}_{\mathrm{I/II}} = 2$~meV) at different temperatures and at various
inter-dot distances. (c) \textit{Left panel:} The long-time asymptotic values of the
spatial coherence for uncoupled QDs as a function of the inter-dot distance. \textit{Right
panel:} The envelope of the oscillations of the asymptotic value of $\rho_{01}$ at larger
distances.}
\end{figure}

\begin{figure}[bt]
\includegraphics[width=85mm]{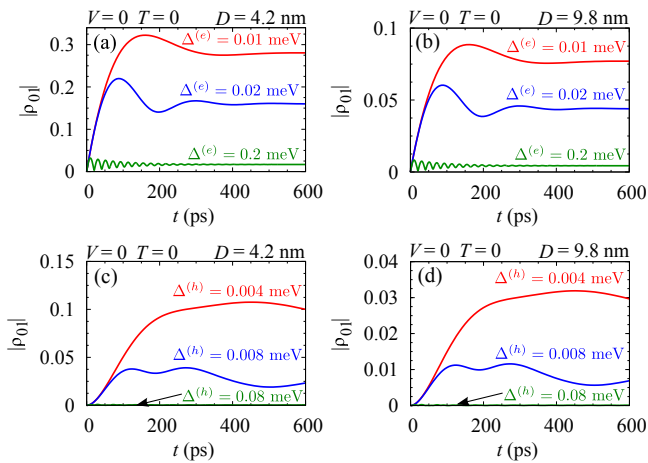}
\caption{\label{fig:deg_noV}(Color online) The evolution of the spatial coherence for
uncoupled QDs at a constant inter-dot distance for a few values of the parameters
characterizing the inhomogeneity of the DQD (denoted $\Delta^{(e)} =
\Delta^{(e)}_{\mathrm{II}} - \Delta^{(e)}_{\mathrm{I}}$ and $\Delta^{(h)} =
\Delta^{(h)}_{\mathrm{II}} - \Delta^{(h)}_{\mathrm{I}}$) for the initial state
$\left(|2\rangle + |3\rangle\right)/\sqrt{2}$ (electron excited) (a,b) and for the initial
state $\left(|6\rangle + |7\rangle\right)/\sqrt{2}$ (electron and hole excited) (c,d). The
values of $\Delta^{(e)}$ in (c,d) are the same as for the corresponding lines in (a,b).}
\end{figure}

The second term in Eq.~(\ref{master_eq2}) describes the transfer of coherence from
$\rho_{23}$ (electron delocalized between the QDs in the excited shell) and $\rho_{45}$
(hole in the excited shell) to the ground state coherence $\rho_{01}$. Let us note that
this term contains the non-secular spectral densities dependent on the inter-dot distance,
as given in Eq.~\eqref{R_D}, hence the resulting transfer of the coherence is sensitive to
the spatial separation of the dots.
\subsubsection*{Results and discussion}
Let us first study the simplest situation, where the system is prepared in the state
$(|2\rangle+|3\rangle)/\sqrt{2}$ (electron at the excited level) in the case of exact
degeneracy ($\omega_{13} = \omega_{02}$, that is, the intraband excitation energies for
the electron are the same in both QDs, $\Delta^{(e)}_{\mathrm{I}} =
\Delta^{(e)}_{\mathrm{II}}$). In this case, the solution of Eq.~\eqref{master_eq2} for the
spatial coherence $\rho_{01}$ at $T = 0$~K reads 
\begin{equation}
\label{ro_01_an}
\rho_{01}=c\Big\lbrack 1-\exp{(-\gamma}t)\Big\rbrack,
\end{equation} where $\gamma = \pi (R_{2002}(\omega_{02})+R_{3113}(\omega_{13})) = 2\pi
R^{(e)}_{\mathrm{sec}}(\omega_{02})$
and
\begin{eqnarray*}
    c
=\frac{1}{2}\frac{R_{3102}(\omega_{02})+R_{1320}(\omega_{13})}{R_{2002}(\omega_{02})+R_{31
13}(\omega_{13})}=\frac{1}{2}\frac{R^{(e)}_{\mathrm{nsec}}(\omega_{02})}{R^{(e)}_{\mathrm{
sec}}(\omega_{02})}.
    \label{nodeg}
\end{eqnarray*}
The results are presented in Fig.~\ref{fig:identical_T0_40} for a few values of the
inter-dot distance. As follows from Eq.~\eqref{ro_01_an}, the rate of the coherence
transfer is directly related to the relaxation rate. However, the maximum value reached by
the spatial coherence depends on the ratio of the magnitude of the additional term in
Eq.~(\ref{master_eq2}) to the usual carrier relaxation (or thermalization) rates, which
reflects the competition between the coherence transfer and its decay in the initial state
due to relaxation. At $T = 0$~K the transitions from the ground state shell to the excited
shells are fully suppressed and the ground state coherence reaches a stable asymptotic
value at long times (Fig.~\ref{fig:identical_T0_40}(a)). Due to the oscillating character
of the non-secular spectral density $R_{\mathrm{nsec}}$, only at certain inter-dot
distances this asymptotic value reaches its maximum which, for flat and not very distant
dots, is very close to 1/2 (see Fig.~\ref{fig:identical_T0_40}(c)). At large distances,
the oscillating factor in Eq.~\eqref{R_D} suppresses the result of the summation, reducing
the coherence transfer rate and the maximum achievable value of $\rho_{01}$. At higher
temperatures the ground state spatial coherence is further suppressed due to thermally
activated processes, which eventually lead to its decay, as shown in
Fig.~\ref{fig:identical_T0_40}(b), where we present the results of numerical simulations
at $T = 40$~K.\\
\indent We have so far assumed the perfect resonance condition which is unlikely in a real
system. The actual importance of the coherence-transfer term in Eq.~\eqref{master_eq2}
depends on the relation between the frequency differences
$\omega_{13}-\omega_{02}=\Delta_{\mathrm{II}}^{(e)}-\Delta_{\mathrm{I}}^{(e)}$ and
$\omega_{15}-\omega_{04}=\Delta_{\mathrm{II}}^{(h)}-\Delta_{\mathrm{I}}^{(h)}$ (see
Fig.~\ref{fig:system}) on one side, and the typical rates of the system evolution in the
interaction picture (coherence transfer and relaxation rates) on the other side.
Therefore, we now proceed to the discussion of a more general case, in which the intraband
energy splittings are not exactly equal.
	\indent This is shown in Fig.~\ref{fig:deg_noV}(a), where we present the evolution of the
ground state spatial coherence for a few values of the energy-level difference between the
dots, for the initial state $\left(|2\rangle + |3\rangle\right)/\sqrt{2}$. We start from
the resonant values ($\Delta^{(e)}_{\mathrm{I/II}} = 5$~meV and
$\Delta^{(h)}_{\mathrm{I/II}} = 2$~meV), then we shift the energy levels away from
resonance in a symmetric way. In Fig.~\ref{fig:deg_noV}(a) we choose the distance $D =
4.2$~nm, for which a local maximum of the non-secular spectral density for the electron is
located at the chosen energy values (blue dashed line and vertical bar in
Fig.~\ref{fig:sd_V0}(a)), while in Fig.~\ref{fig:deg_noV}(b), $D = 9.8$~nm corresponds to
a much lower value of the non-secular spectral density at the relevant energy (gray
dash-dotted line in Fig.~\ref{fig:sd_V0}(a)) and, in consequence, to a lower asymptotic
value of the ground state coherence.
It is clear that even slightly different dots result in a weak spatial coherence. This is
due to the oscillating term in Eq.~\eqref{master_eq} which supresses the transfer of
spatial coherence as soon as the intraband transition energy difference, $\omega_{kl} -
\omega_{mn}$, becomes comparable with the coherence transfer rates. This oscillating
factor leads to oscillations in the ground state spatial coherence in place of the steady
growth appearing at exact resonance. With an increasing inhomogeneity of the QDs, the
oscillations get faster and finally the term responsible for the transfer of coherence
becomes effectively averaged to zero.\\ 
\indent Up to now, our discussion was limited to the direct coherence transfer by a
single-step relaxation (from $\rho_{23}$ to $\rho_{01}$). In the same way, the coherence
$\rho_{01}$ can be fed by $\rho_{45}$. The coherences $\rho_{23}$ and $\rho_{45}$ are
driven by the coherence $\rho_{67}$ in a similar way. Let us now proceed to the general
situation in which the initial state is formed by both the electron and the hole in the
excited states, that is, the fully excited initial state $\left(|6\rangle +
|7\rangle\right)/\sqrt{2}$ (which can be achieved by an optical excitation tuned to the
p-shell of confined states). The evolution of the ground state coherence in this case is
presented in Fig.~\ref{fig:deg_noV}(c,d). Now, the evolution of the spatial coherence
saturates on a longer time scale. This is due to the phonon relaxation of the hole, which
is typically slower than the relaxation of the electron. In addition, large part of the
spatial coherence is lost during hole relaxation, hence the attained asymptotic values are
lower.

\subsection{Coupled QDs}
\label{subsec:coupled}
\subsubsection*{System states}
When the interaction between the dots is included, the system evolution is most
conveniently described in the basis of eigenstates of $H_{\mathrm{DQD}}$. These
single-exciton eigenstates result from the tunnel coupling between the exciton states in
different dots and can be written as
\begin{eqnarray}
|\widetilde{m}\ket & = &  \cos{\frac{\theta_{mn}}{2}}|m\ket +
\sin{\frac{\theta_{mn}}{2}|n\ket},
\label{eigensts1}
\end{eqnarray}
\begin{eqnarray}
|\widetilde{n}\ket & = & -\sin{\frac{\theta_{mn}}{2}}|m\ket +
\cos{\frac{\theta_{mn}}{2}|n\ket}. 
\label{eigensts2}
\end{eqnarray}
Here $m,n$ describe the coupled even and odd states, respectively (that is, a pair of
states with a carrier in different dots but in the same energy shell), $\theta_{mn}$ is
the mixing angle, defined by $\tan\theta_{mn} = 2V_{m}/(\epsilon_{m}-\epsilon_{n})$, where
$\epsilon_{m/n}$ denotes the energy levels in the two dots without the coupling. The
energies of the states $|\widetilde{m}\rangle$ and $|\widetilde{n}\rangle$ are denoted by
$E_{m}$ and $E_{n}$, respectively, where
\begin{equation}
\label{Enm}
E_{n/m}=\frac{\epsilon_{m}+\epsilon_{n}}{2}\pm\frac{1}{2}\Delta E_{nm}
\end{equation}
and 
\begin{equation}
\Delta E_{nm} = \sqrt{(\epsilon_{n}-\epsilon_{m})^{2}+4V^{2}}.\nonumber
\end{equation}
The spatial coherence in the ground state is related to the density matrix elements in the
eigenstate basis by
\begin{eqnarray}
\label{ro_01}
\langle 0|\rho|1\rangle &=&
\frac{1}{2}\sin\theta_{01}\Big(\langle\widetilde{0}|\rho|\widetilde{0}\rangle-\langle\widetilde{1}|\rho|\widetilde{1}\rangle \Big )\nonumber \\
&\quad& +\cos^{2}\frac{\theta_{01}}{2}\langle \widetilde{0}|\rho|\widetilde{1}\rangle
-\sin^{2}\frac{\theta_{01}}{2}\langle \widetilde{1}|\rho|\widetilde{0}\rangle.
\end{eqnarray}

It is clear that the spatial coherence can appear in the ground state as a result of
phonon-induced relaxation even in the absence of the special degeneracy discussed above,
which is a trivial consequence of the fact that the ground state of coupled dots is
coherently delocalized by itself. At $T = 0$~K, when the system relaxes to the ground
state, this coherence is $(1/2)\sin\theta_{01}$. At finite temperatures, the equilibrium
state of the system is
\begin{equation}
\rho = \sum_{n}p_{n}|\tilde{n}\rl \tilde{n}|,
\end{equation}
where $p_{n} =Z^{-1} \exp{(-E_{n}/k_{B}T)}$ and $Z
=\sum_{n}\exp{\Big\lbrack-E_{n}/k_{B}T\Big\rbrack}$. According to Eq.~\eqref{ro_01} this
yields the degree of coherence
\begin{eqnarray}
\label{p0_p1}
\langle 0|\rho|1\rangle &=& \frac{1}{2}\sin\theta_{01}(p_{0} - p_{1}) \\\nonumber
&=&\frac{1}{2}\sin\theta_{01}(p_{0} + p_{1})\tanh\frac{E_{1}-E_{0}}{2k_{B}T}.
\end{eqnarray}
Hence, the coherence at equilibrium depends on the mixing angle (reflecting the relative
coupling strength), the total occupation probability of the ground shell, and the thermal
redistribution of occupations between the two states in the ground shell. In a certain
range of temperatures, when the occupations of the higher shells can be neglected (hence
$p_{0}+p_{1}\approx 1$), the ground state spatial coherence is determined by the last
factor in the above equation.\\
\indent For the numerical simulations of the evolution of the density matrix for coupled
QDs, we use the Master equation as in Eq.~\eqref{master_eq2}, but now in the eigenstate
basis and with the corresponding spectral densities $\widetilde{R}$ (related to those in
the original basis as described in detail in the Appendix B). The energy splittings
$\Delta^{(e)}_{\mathrm{I/II}}$ and $\Delta^{(h)}_{\mathrm{I/II}}$ (see
Fig.~\ref{fig:system}) are obtained by the symmetric displacement of the states with
respect to $\Delta^{(e)}_{\mathrm{I}} = \Delta^{(e)}_{\mathrm{II}} = 5$~meV for the
electron and $\Delta^{(h)}_{\mathrm{I}} = \Delta^{(h)}_{\mathrm{II}} = 2$~meV for the
hole, while the energy difference between the ground states in the dots is constant and
has the value $\Delta^{(e)}_{\mathrm{g}} = -0.05$~meV and $\Delta^{(h)}_{\mathrm{g}} =
-0.008$~meV for the electron and the hole, respectively.
\begin{figure}[bt]
\includegraphics[width=80mm]{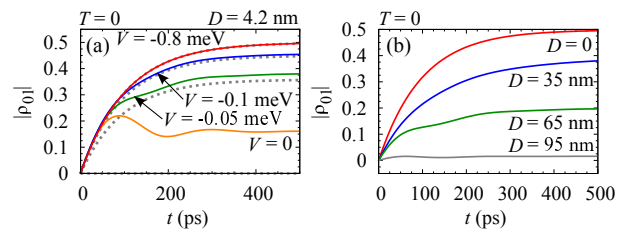}
\caption{\label{fig:LVD}(Color online) The evolution of the spatial coherence for coupled
QDs, with $\Delta^{(e)}=0.02$~meV and $\Delta^{(e)}_{\mathrm{g}} = -0.05$~meV, for the
initial state $\left(|2\rangle + |3\rangle\right)/\sqrt{2}$ (electron excited). (a) For
selected values of the interaction $V$ and a constant inter-dot distance. The gray dotted
lines represent the results obtained from the Lindblad equation (secular approximation)
for the same values of the parameters. (b) For the interaction $V$ that depends on the
distance between the dots.}
\end{figure}
\subsubsection*{Simulation results and discussion}
\begin{figure}[tb]
\includegraphics[width=80mm]{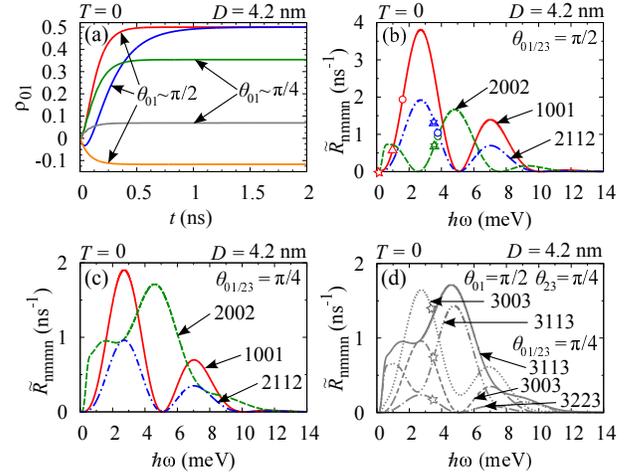}
\caption{\label{fig:fs}(Color online) (a) The evolution of the spatial coherence for
coupled QDs at a constant inter-dot distance. (b,c,d) The phonon spectral densities
describing the electron relaxation within the dots, and the electron transfer between the
states in the dots. The marks in (b,d) (colored circle, triangle and star) denote the
transition energies corresponding to the cases with the red, blue and orange lines in
panel (a), respectively.}
\end{figure}

\begin{figure}[ht]
\includegraphics[width=80mm]{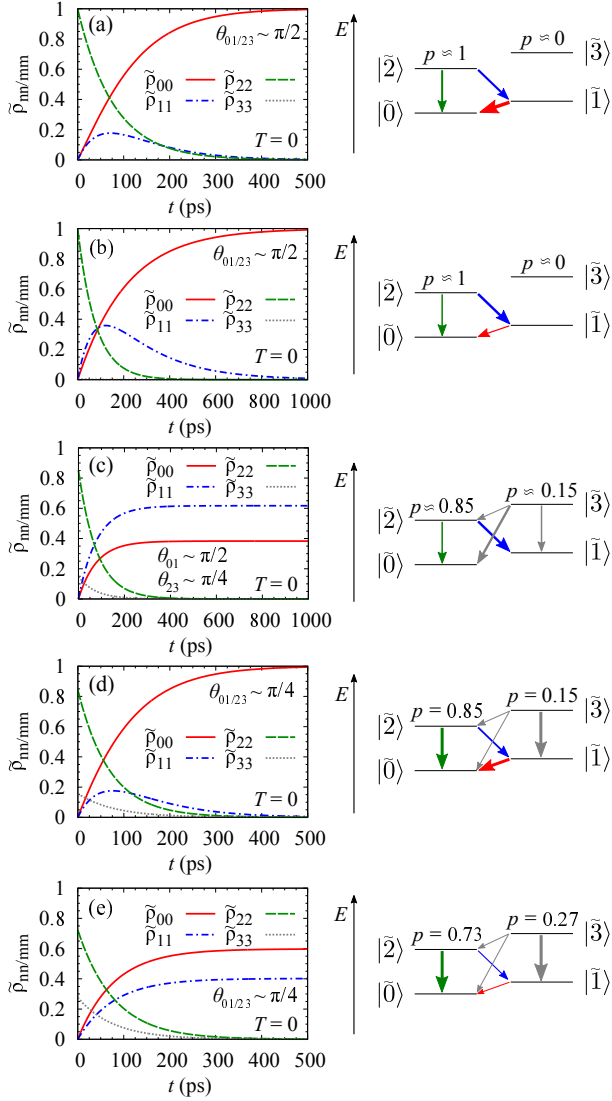}
\caption{\label{fig:occt_combo}(Color online) The occupation of the eigenstates at $T =
0$~K (\textit{left panel}) and the diagram of the transition rates between the states
(\textit{right panel}) for the dynamical scenarios studied in the text. The color of
arrows corresponds to the spectral densities presented in Fig.~\ref{fig:fs}(b,c,d), while
their thickness denotes the efficiency of the relaxation process, and $p$ is the initial
probability of the occupation of the excited states.}
\end{figure}

\begin{figure}[tb]
\includegraphics[width=85mm]{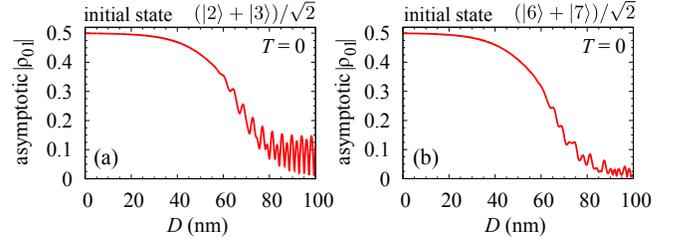}
\caption{\label{fig:ro_01Ve_h}(Color online) The asymptotic values of the spatial
coherence between the two lowest exciton states in the system of coupled and non-identical
QDs at $T = 0$~K. (a) For the initial state $\left(|2\rangle + |3\rangle\right)/\sqrt{2}$
(electron excited), $\Delta^{(e)} = 0.02$~meV. (b) For the initial state $\left(|6\rangle
+ |7\rangle\right)/\sqrt{2}$ (both electron and hole excited), $\Delta^{(e)} = 0.02$~meV,
$\Delta^{(h)} = 0.008$~meV.}
\end{figure}

Again, for the sake of clarity and simplicity, we start our discussion from the case of
the initial state $(|2\rangle+|3\rangle)/\sqrt{2}$ (only electron excited). The results of
these simulations are shown in Fig.~\ref{fig:LVD}. To simplify the discussion, in
Fig.~\ref{fig:LVD}(a) we fixed the inter-dot distance and only increase the coupling. For
the chosen values of the parameters describing the inhomogeneity of the QDs and for a
strong coupling ($V_{m} = V = -0.8$~meV), the simulation results obtained from the
Lindblad equation, that is, in the secular approximation (gray dotted line), are in exact
accordance with the simulation results obtained from the more general Master equation (red
solid line). This means that the evolution of spatial coherence in this regime is
dominated by simple relaxation to the coherently delocalized ground state. For lower
values of the coupling the system reaches the equilibrium value of the coherence, with a
correction to the Lindblad dynamics that grows with decreasing coupling and with some weak
oscillations. These corrections are due to the growing role of the phonon-induced transfer
of coherence from the upper levels. These oscillations do not appear in simulation results
obtained in the secular approximation for the same values of the parameters (gray dotted
lines in Fig.~\ref{fig:LVD}(a)) even though the main trend of the evolution of coherence
as well as the asymptotic value are reproduced correctly unless the coupling becomes very
weak.\\
\indent In Fig.~\ref{fig:LVD}(b) we present the evolution of the spatial coherence in
coupled QDs at $T = 0$~K, when the coupling is related to the distance between the dots.
For this dependence we choose the simplest exponential model\cite{danckwerts06}, $V =
-V_{0}\exp{(-D/D_{0})}$, with the amplitude $V_{0} = 0.5~$meV and the range $D_{0} =
20~$nm. For distant dots (about hundreds of nanometers and more), the coupling vanishes,
hence the equilibrium value of the ground state coherence decreases and the system reaches
the limit of uncoupled dots described in Sec.~\ref{subsec:uncoupled}.  For closely placed
dots, one essentially deals with a relaxation process to the ground state, hence the
ground-state spatial coherence reaches higher values. Weak oscillations visible for
intermediate distances are a fingerprint of the coherence transfer contribution to the
process, which overall remains dominated by relaxation between the eigenstates. Obviously,
the everlasting coherence in our simulation results is due to the lack of exciton decay in
our model and, in a real system, will be limited at longer times by radiative
recombination (see Appendix C for a brief discussion of these effects).\\
\indent In Fig.~\ref{fig:fs}(a) we show the impact of the relaxation within the ground
shell on the system dynamics. Again, we choose the initial state with the excited electron
only, and we fix the inter-dot distance. For better clarity, we keep also constant values
of the intra-dot energy-level differences ($\Delta^{(e)}_{\mathrm{I}} = 3.45$~meV,
$\Delta^{(e)}_{\mathrm{II}} = 3.55$~meV). We consider five scenarios with different
relaxation rates of the electron within and between the dots. The spectral densities
relevant for the discussion of the results in Fig.~\ref{fig:fs}(a) are presented in
Fig.~\ref{fig:fs}(b,c,d) and the associated occupation dynamics for the five cases to be
analyzed is plotted quantitatively and represented diagrammatically in
Fig.~\ref{fig:occt_combo}.\\ 
\indent We start our discussion from strongly coupled QDs with the mixing angle
$\theta_{mn}$ close to $\pi/2$. The spectral densities $\widetilde{R}_{1001}$ and
$\widetilde{R}_{3003}$ (see the Appendix B) involved in the rate of the
$\widetilde{n}\rightarrow\widetilde{m}$ transition at $T = 0$~K ($\widetilde{\Gamma}_{nm}
= 2\pi \widetilde{R}_{nmmn}(\Delta E_{nm}/\hbar)$) for this case are presented in
Fig.~\ref{fig:fs}(b). The initial state in this case is nearly identical with the
eigenstate $|\widetilde{2}\rangle$, which is the lower-energy state of the excited
electron shell. The evolution of the ground state spatial coherence in this case will
depend on the relative values of the transition rates between this state and the two
states in the ground shell of the DQD. Many scenarios are possible, out of which we
discuss the two most characteristic ones. The red line in Fig.~\ref{fig:fs}(a)
($\Delta^{(e)}_{g} = -0.05$~meV, $V = -0.8$~meV, $\theta_{01} = 88.21^{\circ}$,
$\theta_{23} =  84.64^{\circ}$) represents the evolution of the ground-state spatial
coherence with fast relaxation  between the two states in the ground shell (case (a) in
Fig.~\ref{fig:occt_combo}). In this case, the occupation of the ground states changes fast
and the spatial coherence reaches its asymptotic value in a short time. During the
evolution one can observe only a small transient occupation of the state
$|\widetilde{1}\rangle$ (see Fig.~\ref{fig:occt_combo}(a)).
\\
\indent The blue line in Fig.~\ref{fig:fs}(a) ($\Delta^{(e)}_{g} = -0.03$~meV, $V =
-0.5$~meV, $\theta_{01} = 88.17^{\circ}$, $\theta_{23} = 82.48^{\circ}$) corresponds to
the situation where the efficiency of the carrier transfer between the ground states is
lower as compared to the relaxation to the state $|\widetilde{1}\rangle$ (see
Fig.~\ref{fig:occt_combo}(b)). Although the system still finally relaxes to the
delocalized ground state $|\widetilde{0}\rangle$ with the maximum value of the spatial
coherence of 1/2, the final ground-state spatial coherence saturates after a much longer
time. \\
\indent Now, we focus on weakly coupled QDs, where initially both of the excited-shell
eigenstates are occupied. In such a  situation, the efficiency of the carrier transfer
between the dots strongly affects the value of the ground-state spatial coherence. In the
limiting case of vanishing relaxation within the ground shell, the system state at $T =
0$~K is asymptotically frozen in a mixture of the state $|\widetilde{0}\rangle$ and
$|\widetilde{1}\rangle$, with relative probabilities determined by the relaxation rates
from $|\widetilde{3}\rangle$ and $|\widetilde{2}\rangle$ to these two eigenstates. Since
the ``anti-bonding'' state $|\widetilde{1}\rangle$ contributes negative spatial coherence,
the resulting value of this coherence may in principle achieve any value between -1/2 and
1/2. An example is shown by the orange line in Fig.~\ref{fig:fs}(a) ($\Delta^{(e)}_{g} =
-0.001$~meV, $V = -0.05$~meV, $\theta_{01} = 89.43^{\circ}$, $\theta_{23} =
44.71^{\circ}$) and in Fig.~\ref{fig:occt_combo}(c). At sufficiently high temperatures,
this state would become meta-stable and decay to the ground state $|\widetilde{0}\rangle$
due to thermally activated transitions via the excited states. \\
\indent The evolution in the case of weakly coupled QDs is shown in Fig.~\ref{fig:fs}(a)
by the green line (fast carrier transfer between the dots, $\Delta^{(e)}_{g} = -1.6$~meV,
$V = -0.8$~meV, $\theta_{01} = 45^{\circ}$, $\theta_{23} = 43.26^{\circ}$) and gray line
(slow transfer, $\Delta^{(e)}_{g} = -0.1$~meV, $V = -0.05$~meV, $\theta_{01} =
45^{\circ}$, $\theta_{23} = 26.57^{\circ}$). The spectral densities relevant to these
cases are shown in Fig.~\ref{fig:fs}(c,d). In the case of fast transfer, the electron
transfer between the states in the ground shell is relatively efficient (see
Fig.~\ref{fig:occt_combo}(d)) and the final ground-state spatial coherence quickly reaches
its asymptotic value as in Eq.~\eqref{p0_p1}. In the other case, when the transfer within
the ground shell is extremely slow (see Fig.~\ref{fig:occt_combo}(e)), the final degree of
coherence results from the coherence transfer process, with the final value yielded by the
interplay of the occupation relaxation and coherence transfer rates, as discussed in the
previous section. In the particular case shown in Fig.~\ref{fig:fs}(a) (gray line), this
value is lower than that associated with the ground state (green line).\\
\indent In Fig.~\ref{fig:ro_01Ve_h} we present the long-time asymptotic values of the
spatial coherence in coupled QDs at $T = 0$~K, as a function of the inter-dot distance.
For small distances, the effect is dominated by the ground state coherence. Since the
mixing angle decreases with vanishing coupling, so does the amount of spatial coherence in
the ground state and, as a result, the final degree of spatial coherence decreases with
the inter-dot distance. For larger distances between the dots the efficiency of the
coherence transfer is most important (uncoupled dots limit, see
Sec.~\ref{subsec:uncoupled}), hence the asymptotic value oscillates with the distance, in
view of the corresponding form of the spectral density (as in
Fig.~\ref{fig:identical_T0_40}(c), but with a smaller amplitude because of non-identical
dots). Furthermore, the coupling becomes smaller than the energy mismatch and the
resulting maximum value of the coherence is reduced. The resulting dependence on the
distance is similar for the initially excited electron (Fig.~\ref{fig:ro_01Ve_h}(a)) and
for both the electron and the hole excited (Fig.~\ref{fig:ro_01Ve_h}(b)).\\
\begin{figure}[tb]
\includegraphics[width=85mm]{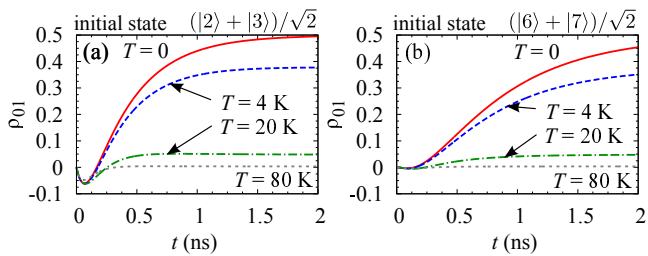}
\caption{\label{fig:ro_01V}(Color online) The evolution of the spatial coherence between
the two lowest exciton states in the system of non-identical QDs at various temperatures.
(a) $\Delta^{(e)} = 0.02$~meV. (b) $\Delta^{(e)} = 0.02$~meV, $\Delta^{(h)} = 0.008$~meV.}
\end{figure}
\indent An important feature of phonon-related effects is their strong temperature
dependence. In Fig.~\ref{fig:ro_01V} we show the evolution of the spatial coherence in
coupled QDs at a few values of the temperature. In Fig.~\ref{fig:ro_01V}(a) we present the
results for the initial state $\left(|2\rangle + |3\rangle\right)/\sqrt{2}$ (excited
electron). In Fig.~\ref{fig:ro_01V}(b), results for the initial state $\left(|6\rangle +
|7\rangle\right)/\sqrt{2}$ are presented where, at the beginning, the electron and hole
are in their excited states. The thermal distribution of the occupations reduces the
asymptotic degree of coherence in accordance with Eq.~\eqref{p0_p1}. At a low temperature
(blue dashed line), mostly the ground states are occupied. At higher temperatures the
occupations are distributed between the ground and excited states, hence the ground state
coherence is suppressed. Let us note that in comparison with the electron, the hole has a
lower value of the deformation potential constant. Hence, the phonon relaxation of the
hole is slower and the resulting evolution of the spatial coherence in the case of both
carriers initially excited saturates after a longer time.
\section{Discussion and conclusions}
\label{sec:conclusions}
We have studied theoretically the evolution of the spatial coherence during relaxation
between delocalized exciton states in the system consisting of two vertically stacked
semiconductor QDs coupled to the acoustic phonons. We have shown that the coupling to
phonons can lead to coherence transfer from a delocalized excited state to the ground
state manifold of the DQD, which accompanies phonon-induced relaxation in this system. As
we have found, the coherence transfer in a system of decoupled dots requires that the
intraband energy spacings in the two dots should be close to each other. In the presence
of coupling between the dots, some spatial coherence is always present in the system
ground state. Hence, in the evolution, one deals with an interplay between the coherence
transfer process (which is related to intra-dot relaxation), and the relaxation to the
ground state (which involves inter-dot transfer, enabled by the coupling).\\
\indent As a final element of our discussion, let us see to what extent the required
degeneracy of intraband transition energies is likely to appear in a real system. The
inhomogeneous broadening of ground state photoluminescence lines in QD ensembles is
typically on the order of $10$~meV. Apart from being composed of electron and hole
contributions, this broadening is mostly due to the fluctuation in the QD size $H$ along
the strongest confinement axis (i.e., the growth direction). Since the energy is
proportional to inverse square of the size, the variation of the intraband transition
energies, which depend on the lateral size $L$, is reduced by a factor $(H/L)^2$, that is,
by one or two orders of magnitude. This estimate yields a typical inhomogeneity of
electron intraband transition energies of $0.1$~meV and a correspondingly smaller value
for the hole. These energies may lie even closer to each other in particular QD pairs. On
the other hand, the critical value of the inhomogeneity that destroys the coherence
transfer effect is related to the onset of validity of the secular approximation and hence
depends on the typical relaxation rate. In smaller QDs, where the LO-phonon mediated
relaxation is faster by an order of magnitude than in our present model, the tolerance of
this effect against inhomogeneity will also be enhanced by an order of magnitude.\\
\indent The present study was based on a minimal model with the aim to demonstrate that
the process in question is possible in principle and to characterize the most general
conditions that must be met for this process to appear. Obviously, it must be supplied
with much more details concerning the actual system spectrum and the full set of phonon
modes and processes that contribute to relaxation in a typical structure (most likely
optical phonons and anharmonicity-related processes) in order to yield quantitative
conclusions on the importance of the coherence transfer process under particular
experimental conditions. In the detailed model, one might also include the exact structure
of wave functions. In particular, the product form of the exciton wave functions is not
strictly consistent with the assumption that the Coulomb attraction is larger than the
differences between single particle levels. Including Coulomb-correlated exciton states
would yield more accurate values of transition rates, although the general structure of
the model and, therefore, the essential conclusions of our analysis would remain
unaffected. Apart from the possible quantitative improvements of the model, already at the
present stage of a ``proof of principle'', our results indicate that the interplay of the
coupling and energy degeneracy is crucial not only for the appearance of collective
emission itself \cite{karwat11,sitek07a} but also for the coherence transfer during
relaxation that is needed for the appearance of the coherent Dicke-like state in the
ground shell of the DQD. 

\acknowledgments
This work was supported by the grant No. 2011/01/B/ST3/02415 from the Polish National
Science Centre.
\setcounter{secnumdepth}{0}

\section{Appendix A: Non-secular master equation preserves physicality} 
\label{sec:appa}

In this Appendix we show that retaining the non-secular terms in
Eq.~\eqref{master_eq} does not break the physicality of the model in the sense that
the density matrix remains positive-definite during the evolution
generated by this master equation. We do this by explicitly
transforming Eq.~\eqref{spectral_density} into the Lindblad form, which is known to posses
the required properties \cite{breuer02}.  

For the sake of clarity, let us define $\sigma_{mn}=|m\rl n|$. We
start by rewriting Eq.~\eqref{spectral_density} in terms of the density matrix in the
Schr\"odinger picture, $\rho_{\mathrm{S}}=e^{-iH_{0}t/\hbar}\rho
e^{iH_{0}t/\hbar}$, where $H_{0}=\sum_{n}\epsilon_{n}|n\rl n|$. One
finds 
\begin{displaymath}
\dot{\rho_{\mathrm{S}}}= -\frac{i}{\hbar}\left[ H_{0},\rho_{\mathrm{S}} \right] 
+\mathcal{L}_{\mathrm{S}}[\rho_{\mathrm{S}}],
\end{displaymath}
where
\begin{eqnarray*}
\lefteqn{\mathcal{L}_{\mathrm{S}}[\rho_{\mathrm{S}}] =}  \\
&&\pi \sum_{lkmn} R_{lkmn}(\omega_{mn}) \left(
\sigma_{mn}\rho_{\mathrm{S}}\sigma^{\dag}_{kl} 
- \sigma^{\dag}_{kl}\sigma_{mn}\rho_{\mathrm{S}} \right)
+\mathrm{h.c.} 
\end{eqnarray*}
By appropriately renaming the summation indices and using the symmetry
relations for the spectral densities listed below Eq.~\eqref{spectral_density}, the
dissipator can be written in the form
\begin{displaymath}
\mathcal{L}_{\mathrm{S}}[\rho_{\mathrm{S}}] =
-\frac{i}{\hbar}\left[  H_{1},\rho_{\mathrm{S}} \right] 
+\mathcal{L}'[\rho_{\mathrm{S}}],
\end{displaymath}
where
\begin{displaymath}
\mathcal{L}'[\rho_{\mathrm{S}}] = 
2\pi \sum_{klmn} \Gamma_{(kl)(mn)} \left( \sigma_{mn}\rho_{\mathrm{S}}\sigma^{\dag}_{kl} 
- \left\{ \sigma^{\dag}_{kl}\sigma_{mn},\rho_{\mathrm{S}} \right\} \right),
\end{displaymath}
with 
$\Gamma_{(kl)(mn)}=[R_{lkmn}(\omega_{mn})+R_{lkmn}(\omega_{kl})]/2$,
and
\begin{displaymath}
H_{1}=\frac{i}{2}\sum_{klmn} 
\left[ R_{lkmn}(\omega_{mn})-R_{lkmn}(\omega_{kl}) \right] 
\sigma_{kl}^{\dag}\sigma_{mn}.
\end{displaymath}

In view of the symmetries of the spectral densities, $H_{1}$ is
a hermitian operator and the matrix of transition rates $\Gamma$ is also hermitian, that
is, $\Gamma_{(kl)(mn)}=\Gamma^{*}_{(mn)(kl)}$. The dissipator
$\mathcal{L}'$ can therefore be brought to the standard Lindblad form
\cite{breuer02}. One diagonalizes $\Gamma$ by a unitary matrix $U$,
\begin{displaymath}
\sum_{klmn}U_{(kl),\alpha}^{*} \Gamma_{(kl)(mn)}U_{(mn),\beta} 
= \gamma_{\alpha}\delta_{\alpha\beta},
\end{displaymath}
with $\sum_{kl} U_{(kl),\alpha} U_{(kl),\beta}^{*}
=\delta_{\alpha\beta}$ and $\sum_{\alpha} U_{(kl),\alpha} U_{(k'l'),\alpha}^{*}
=\delta_{kk'}\delta_{ll'}$. Define
$\sigma_{\alpha}=\sum_{kl}U^{*}_{(kl),\alpha}\sigma_{kl}$, that is,
$\sigma_{kl} =\sum_{\alpha}U_{(kl)\alpha}\sigma_{\alpha}$. In terms
of these new operators one finds
\begin{displaymath}
\mathcal{L}'[\rho_{\mathrm{S}}]=\sum_{\alpha}\widetilde{\Gamma}_{\alpha}\left[ 
\sigma_{\alpha} \rho_{\mathrm{S}} \sigma_{\alpha}^{\dagger}
-\frac{1}{2}\left\{ \sigma_{\alpha}^{\dag}\sigma_{\alpha},\rho_{\mathrm{S}} \right\} 
 \right].
\end{displaymath}
Hence, Eq.~\eqref{spectral_density} in the Schr\"odinger picture has the form of the
Lindblad equation
\begin{displaymath}
\dot{\rho_{\mathrm{S}}}= -\frac{i}{\hbar}\left[ H_{0}+H_{1},\rho_{\mathrm{S}} \right] 
+\mathcal{L}'[\rho_{\mathrm{S}}]
\end{displaymath}
and therefore preserves the physicality of the density matrix.

\section{Appendix B: Spectral densities for coupled QDs} 
\label{sec:appb}
In this appendix, we summarize the calculation of the spectral densities for a coupled
system, which are related to those in the original basis, defined in
Sec.~\ref{subsec:uncoupled}. We focus on the secular and non-secular spectral densities
for the electron relaxation. The same calculations can be repeated for the hole
relaxation.\\
\indent Upon transformation to the eigenbasis defined by Eqs.~\eqref{eigensts1} and
\eqref{eigensts2}, Eq.~\eqref{ham-phon} becomes
\begin{eqnarray}
\label{ham-phon-inter}
H_{\mathrm{c-ph}}&=&\sum_{(mn)}|\widetilde{m}\rl \widetilde{n}|\sumk
\widetilde{F}_{mn}(\bm{k})(\bk+\bmkd)\\&+& \sum_{(n'n)}|\widetilde{n}'\rl
\widetilde{n}|\sumk \widetilde{F}_{n'n}(\bm{k})(\bk+\bmkd).\nonumber
\end{eqnarray}
The first term in the Hamiltonian is derived from the first term in Eq.~\eqref{ham-phon}
and corresponds to inter-shell transitions. The coupling between the QDs and the resulting
state mixing opens new (cross-QD) relaxation channels, compared to Eq.~\eqref{ham-phon},
where only relaxation within a QD was possible. This is accounted for by a larger set of
coupling constants, e.g.,

\begin{equation}
\widetilde{F}_{02}(\bm{k}) =
\cos{\frac{\theta_{01}}{2}}\cos{\frac{\theta_{23}}{2}}F_{02}(\bm{k})+\sin{\frac{\theta_{01
}}{2}}\sin{\frac{\theta_{23}}{2}}F_{13}(\bm{k}).\nonumber
\end{equation}
The second term, in which $[n'n]$ denotes summation over pairs of states belonging to one
shell, originates from the diagonal couplings which did not contribute to the dynamics in
the uncoupled case in the Markov limit. In the presence of the coupling between the QDs,
these diagonal terms yield transitions between the delocalized states that form a doublet
within a given shell. For instance,
\begin{equation}
\widetilde{F}_{01}(\bm{k}) =
\frac{1}{2}\sin\theta_{01}(F_{11}(\bm{k})-F_{00}(\bm{k})).\nonumber
\end{equation}
 The spectral densities for the coupled case are then calculated according to
Eq.~\eqref{spectral_density} using the transformed coupling constants. As a result, one
obtains, e.g., for the spectral densities most relevant to the discussion in
Sec.~\ref{subsec:coupled},
\begin{eqnarray*}
    \widetilde{\mathrm{R}}_{2002}(\omega) = \widetilde{\mathrm{R}}_{3113}(\omega) &=&
C_{1}\mathrm{R}^{(e)}_{\mathrm{sec}}(\omega)+
C_{2}\mathrm{R}^{(e)}_{\mathrm{nsec}}(\omega),
\end{eqnarray*}
\begin{eqnarray*}
    \widetilde{\mathrm{R}}_{2112}(\omega) = \widetilde{\mathrm{R}}_{3003}(\omega) &=&
C_{3}\mathrm{R}^{(e)}_{\mathrm{sec}}(\omega) -
C_{2}\mathrm{R}^{(e)}_{\mathrm{nsec}}(\omega),
\end{eqnarray*}
\begin{eqnarray*}
\lefteqn{\widetilde{\mathrm{R}}_{1001}(\omega) = \frac{1}{4}\sin^{2}\theta_{01}
\Big[\mathrm{R}^{(e)}_{\mathrm{0000}}(\omega)+\mathrm{R}^{(e)}_{\mathrm{1111}}(\omega)}\\
&\quad&-\mathrm{R}^{(e)}_{\mathrm{1100}}(\omega)-\mathrm{R}^{(e)}_{\mathrm{0011}}(\omega)\Big], 
\end{eqnarray*}
where 
\begin{eqnarray*}
C_{1} &=& \sin^{2}\frac{\theta_{01}}{2}\sin^{2}\frac{\theta_{23}}{2} +
\cos^{2}\frac{\theta_{01}}{2}\cos^{2}\frac{\theta_{23}}{2},\\
C_{2} &=& \frac{1}{2}\sin\theta_{01}\sin\theta_{23},\\
C_{3} &=& \sin^{2}\frac{\theta_{01}}{2}\cos^{2}\frac{\theta_{23}}{2} +
\cos^{2}\frac{\theta_{01}}{2}\sin^{2}\frac{\theta_{23}}{2}.\\
\end{eqnarray*} 

\section{Appendix C: The effect of finite exciton life time} 
\label{sec:appb}

Spontaneous emission (radiative recombination of excitons) is modeled by the Lindblad
equation, written in the basis of exciton eigenstates $|\widetilde{l}\rangle$
[Eqs.~\eqref{eigensts1} and \eqref{eigensts2}],
\begin{eqnarray*}
\lefteqn{\mathcal{L}_{\mathrm{rad}}[\rho] =}  \\
&& \sum_{l} \gamma_{l} \left[ \Sigma_{l}\rho\Sigma^{\dag}_{l} 
- \frac{1}{2}\left(\Sigma^{\dag}_{l}\Sigma_{l}\rho+\rho\Sigma^{\dag}_{l}\Sigma_{l}
\right)\right],
\end{eqnarray*}
where $\Sigma_{l}$ annihilates the exciton in the state $|l\rangle$ and $\gamma_{l}$ is
the recombination rate for this state. The latter depend on the delocalization of the
exciton state \cite{sitek07a,karwat11}. Assuming that the electron and hole wave functions
in a given dot are similar and that the excited  states $|6\rangle$ and $|7\rangle$ are
bright, one has for the states defined in Eqs.~\eqref{eigensts1} and \eqref{eigensts2}
$\gamma_{0,1} = \gamma_0 (1\pm \sin\theta_{01})$, $\gamma_{6,7} = \gamma_0 (1\pm
\sin\theta_{67})$ (the upper sign corresponds to the first of the two indices), and all
the remaining decay rates equal to 0. Here $\gamma_0$ is the decay rate of an exciton on a
single quantum dots; we assume $\gamma_0 = 1$~ns$^{-1}$.
\begin{figure}[bt]
\includegraphics[width=80mm]{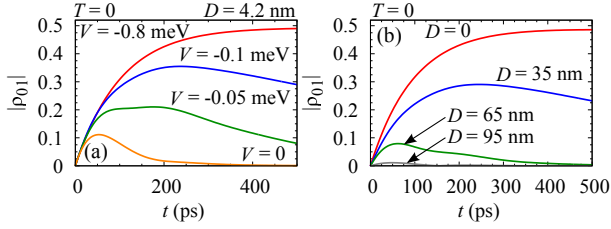}
\caption{\label{fig:LVD2n}(Color online) The evolution of the spatial coherence for
coupled QDs (including the radiative recombination of exciton), with the same values of
the parameters $\Delta^{(e)}$ and $\Delta^{(e)}_{\mathrm{g}}$ as in Fig.~\ref{fig:LVD},
again for the initial state with the excited electron only ($\left(|2\rangle +
|3\rangle\right)/\sqrt{2}$). (a) For selected values of the interaction $V$ and a constant
inter-dot distance. (b) For the interaction $V$ that depends on the distance between the
dots.}
\end{figure}\\
\indent As an example, we recalculate the results presented in Fig.~\ref{fig:LVD}, now
taking into account the radiative decay. The results are shown in Fig.~\ref{fig:LVD2n}.
The effect at short times is very small. On the long time scales, however, the coherence
does not stabilize at a constant value but decays due to radiative recombination of the
exciton. As a result of this process, the maximum value of the spatial coherence is lower
than in the idealized model discussed in this paper.

\bibliographystyle{prsty}
\bibliography{abbr,quantum}

\end{document}